\documentclass[%
 aip,
 amsmath,amssymb,
preprint,%
]{revtex4-1}

\usepackage{graphicx}
\usepackage{dcolumn}
\usepackage{bm}

\usepackage[T1]{fontenc}
\usepackage{mathptmx}
\usepackage{braket}
\usepackage{amsmath}

\newcommand\tr{\operatorname{tr}}
\newcommand\dscf{\(\Delta\)SCF}
\usepackage{xcolor}

\UseRawInputEncoding

\begin{document}

\author{Susannah Bourne Worster}
\email[]{susannah.bourne-worster@bristol.ac.uk}
\affiliation{Centre for Computational Chemistry, School of Chemistry, University of Bristol, Bristol BS8 1TS, U.K.}

\author{Oliver Feighan}
\affiliation{Centre for Computational Chemistry, School of Chemistry, University of Bristol, Bristol BS8 1TS, U.K.}

\author{Frederick R. Manby}
\email[]{fred.manby@bristol.ac.uk}
\altaffiliation{Present address: Entos, Inc., 4470 W Sunset Blvd Suite 107 PMB 94758, Los Angeles, CA 90027, U.S.A.}
\affiliation{Centre for Computational Chemistry, School of Chemistry, University of Bristol, Bristol BS8 1TS, U.K.}

\date{\today}

\begin{abstract}
Delta-self-consistent field theory (\dscf{}) is a conceptually simple and computationally inexpensive method for finding excited states. Using the maximum overlap method to guide optimization of the excited state, \dscf{} has been shown to predict excitation energies with a level of accuracy that is competitive with, and sometimes better than, that of TDDFT. Here we benchmark \dscf{} on a larger set of molecules than has previously been considered, and, in particular, we examine the performance of \dscf{} in predicting transition dipole moments, the essential quantity for spectral intensities. A potential downfall for \dscf{} transition dipoles is origin dependence induced by the nonorthogonality of \dscf{} ground and excited states. We propose and test the simplest correction for this problem, based on symmetric orthogonalization of the states, and demonstrate its use on bacteriochlorophyll structures sampled from the photosynthetic antenna in purple bacteria.
\end{abstract}

\title{Reliable transition properties from excited-state mean-field calculations}

\maketitle

\section{Introduction}

The matrix element of the electric dipole operator \(\hat{\bm{\mu}}\) between two quantum states, commonly known as a transition dipole moment \(\bm{\mu}\), is a crucial quantity in simulating spectra and describing excited-state dynamics of molecular systems. The magnitude of the transition dipole moment \(|\bm{\mu}|\) defines the strength with which a transition between the two states can couple to the electromagnetic field to absorb (or emit) light, while the dipole-dipole interaction between transition dipole moments provides the simplest model for the coupling between excited states on different chromophores.

An important application of this second property is in describing transport of excitons through a network of chromophores, as is seen in the early stages of photosynthesis, as well as synthetic analogues, such as organic polymer light-emitting diodes \cite{Barford2013b} and chromophores hosted on DNA scaffolds.\cite{Buckhout-White2014,Hemmig2016,Olejko2017}  These systems are often simulated using a Frenkel exciton Hamiltonian, \cite{Tretiak2000,Jordanides2002,Manzano2013,BourneWorster2019}
\begin{equation}
    \hat{H} = \sum_i E_i\ket{i}\!\bra{i} + \sum_{i\ne j}V_{ij}\ket{i}\!\bra{j},
\end{equation}
whose off-diagonal elements, \(V_{ij}\), are the coulomb interaction between the transition dipole moments of the relevant excitation on each chromophore. The light-harvesting antenna in photosynthetic organisms typically contain large numbers of chromophores, which are, themselves, relatively large conjugated organic molecules. For example, the antenna in purple photosynthetic bacteria consists of 3--10 light-harvesting II (LHII) complexes (and one LHI complex) per reaction centre, \cite{Trissl1999,Scheuring2004} each containing 27 (32) bacteriochlorophyll-a (BChla) chromophores of around 140 atoms.\cite{Cherezov2006,Hu2002a} Furthermore, the transition dipole moment of each chromophore, and hence the coupling elements of the Hamiltonian, fluctuate constantly with the vibrations of the molecules. To capture the full time-dependent Hamiltonian, even approximately, calculation of the transition dipole moments should therefore ideally be computationally cheap, as well as reasonably accurate. Current models of exciton dynamics in these systems rely on parameterising the coupling elements \(V_{ij}\) from experiment, \cite{Tretiak2000,Ishizaki2009,Mohseni2013,Wu2010,Pullerits1996,Pelzer2014} or use time-dependent density functional theory (TDDFT) to generate representative transition dipole moments from a small handful of chromophores.\cite{Stross2016,Beljonne2009} On-the-fly TDDFT has been used in this context for a single LHII complex, \cite{Sisto2017} and the present work forms part of a wider effort to scale and refine the approach reported there.

TDDFT \cite{Runge1984,Casida1995,Stratmann1998} is a widely popular method for obtaining the properties, including transition dipole moments, of excited states.
\cite{Adamo2013,Laurent2013} With the right choice of exchange-correlation functional and basis set, it yields good accuracy  compared to correlated wavefunction methods such as CC2 \cite{Sarkar2020} and EOM-CCSD \cite{Caricato2011,Laurent2013}, at a much lower computational cost.  In TDDFT, excitation energies  emerge as the eigenvalues of the Casida equations.\cite{Casida1995,Furche2005a}

The transition vectors (which arise as eigenvectors) are expressed in a basis of excitations \(i\rightarrow a\) and corresponding de-excitations. Each one can be reshaped into a transition density matrix, with columns \(i\) and rows \(a\), from which the transition properties of the excited state can easily be calculated.  The transition dipole moment, for example, is found by tracing the transition density matrix with the dipole operator \(\hat{\boldsymbol\mu}\).

However, TDDFT is still too costly  to perform dynamics calculations involving large numbers of BChla chromophores, and this paper amounts to an investigation into how feasible it  would be to use the cheaper \dscf{} method. 
Crucially \dscf{} is not only simpler for the energy evaluation; the excited-state gradient is also available very cheaply because it can be computed using standard ground-state mean-field gradient theory.

\dscf{} is conceptually simple.  Excited states are found by promoting an electron from an occupied orbital in the ground state to one of the unoccupied virtual orbitals.  The orbitals are then reoptimised for the excited electron configuration using a normal SCF  iterative procedure. \cite{Hunt1969,Huzinaga1970,Huzinaga1971,Morokuma1972,Gilbert2008} Unlike TDDFT, therefore, \dscf{} produces a distinct set of molecular orbitals for the excited state.  The transition dipole moment can be calculated as a matrix element between the ground-state and excited determinants.  

Initial attempts to locate excited states via an SCF procedure rigidly maintained the orthogonality of the ground and excited states by relaxing the excited state particle (and hole) orbitals within the ground-state virtual space \cite{Hunt1969,Huzinaga1970} (or respectively in the virtual and occupied spaces\cite{Morokuma1972}). In addition to the convenience of dealing with orthogonal states, these procedures also ensure that relaxing the orbitals does not collapse the excited state wavefunction back down to the ground state. Gilbert {\it et al.} later argued that imposing orthogonality in this way led to wavefunctions that were no longer solutions of the full SCF equations and propagated errors and approximations in the ground state. \cite{Gilbert2008}  They relaxed the orthogonality condition and searched for high energy solutions to the SCF equations by minimising the energy of the excited state with the added condition that the occupied orbitals at each step of the iterative cycle should overlap as much as possible with their counterparts in the previous iteration.  This is known as the maximum overlap method (MOM) and has been shown to be highly successful in finding excited state energies. \cite{Gilbert2008,Barca2018,Hanson-Heine2013,Kowalczyk2011} The sizeable test set that we consider in this paper adds to this body of evidence, as well as benchmarking the technique for transition dipole moments.

However, as Gilbert {\it et al.} acknowledge in their original paper, allowing non-zero overlap between the ground and excited state can artificially enhance the size of the transition dipole moment (and other transition properties).
Nonorthogonality of the states introduces a non-zero transition charge, equal to the size of the overlap. Transition dipole moments calculated from the charged transition density are origin-dependent and therefore have a completely arbitrary magnitude. When the state overlap is very small and the molecule is is positioned with its centre of mass on, or close to, the origin, the error associated with the charged transition density is small, or even negligible. Conversely, if the molecule is positioned far away from the origin, as might be the case for a chromophore located within a larger complex or aggregate centred collectively on the origin, the error associated with this additional charge can quickly escalate.  Here we propose and test a simple correction that can be applied to the transition density matrix after the SCF cycle, to restore the orthogonality of the ground and excited state.

\section{Theory}

The transition dipole for an excitation from an initial state \(\ket{\Psi_1}\) to a final state \(\ket{\Psi_2}\) is defined in the standard length gauge as

\begin{equation}
    \boldsymbol\mu^{1\rightarrow 2} = \braket{\Psi_2|\hat{\boldsymbol\mu}|\Psi_1},\label{eq:def_trans_dip}
\end{equation}
where \(\hat{\boldsymbol\mu}\) is the 3-component dipole operator.


In \dscf\ the states $\ket{\Psi_n}$ are Slater determinants constructed from spin orbitals $\{\ket{\phi^{(n)}_j}\}$ with $n=1,2$. The orbitals are orthonormal within each state, but, in general, nonorthogonal between states, with inner products  $S^{21}_{jk}=\langle\phi_j^{(2)}|\phi_k^{(1)}\rangle$.
%
%
%
The inner product of the two determinants is the determinant of the orbital inner products: 
\begin{equation}
    \braket{\Psi_2|\Psi_1} = |\mathbf S^{21}| \;.
\end{equation}


Following the normal rules for nonorthogonal determinants laid down by L\"owdin \cite{Lowdin1955}, the transition dipole can be written
\begin{equation}
    \braket{\Psi_2|\hat{\boldsymbol\mu}|\Psi_1}
  =
  \sum_{jk} \boldsymbol\mu^{21}_{jk} \operatorname{adj}(\mathbf S^{21})_{jk}
  \;,
\end{equation}
where $\boldsymbol\mu^{21}_{jk}=\braket{\phi^{(2)}_j|\hat{\boldsymbol\mu}|\phi^{(1)}_k}$ and where  $\operatorname{adj}$ denotes matrix adjugate.

%
Alternatively the value of the transition dipole can be computed from the reduced one-particle transition density matrix:
\begin{equation}
    \braket{\Psi_2|\hat{\boldsymbol\mu}|\Psi_1} 
    = \tr(\hat{\boldsymbol\mu}\ket{\Psi_1}\!\bra{\Psi_2}) 
    = \tr(\boldsymbol\mu\mathbf D^{21}).
    \label{eq:td_as_trace_with_tdm}
\end{equation}
Here $\mathbf D^{21}$ is the one-particle reduced transition density matrix in the atomic-orbital basis, given by
\begin{equation}
    \mathbf D^{21} = \mathbf C^{(2)}
    \operatorname{adj}(\mathbf S^{21})
    \mathbf C^{(1)\dagger}
\end{equation}
where $\mathbf C^{(n)}$ are the molecular-orbital coefficients for state $n$. 
%
%
For unrestricted calculations the spin summation for the reduced density matrix has additional factors that would be 1 or 0 if a common set of orthonormal orbitals were being used, but here have to be considered explicitly:
%
\begin{equation}
    \mathbf D^{21} = 
    \mathbf D^{21,\alpha}|\mathbf S^{21,\beta}|
    +
    \mathbf D^{21,\beta}|\mathbf S^{21,\alpha}| \;,
\end{equation}
where 
$\mathbf D^{21,\sigma}$ is the analogue of $\mathbf D^{21}$ for the $\sigma$ spin channel.

As noted above, in \dscf{}, the sets of orbitals \(\{\phi^{(1)}_j\}\) and \(\{\phi^{(2)}_k\}\) for the ground and excited states are optimised independently, so that the resulting states \(\ket{\Psi_1}\) and \(\ket{\Psi_2}\) are not necessarily orthogonal. As previously recognised in the literature,\cite{Gilbert2008}  non-zero overlap between these two states leads to errors in the calculated transition dipole moment. In particular, when states are not exactly orthogonal there is a non-zero transition charge equal to the value of the overlap:
$q^{21} = \braket{\Psi_2|\Psi_1}$. 
This breaks the origin-independence of  the transition dipole moment, making the calculated values virtually meaningless. While the transition charge is sometimes exactly zero (when the ground and excited states are of different symmetries) or very small, any violation of translational invariance is certain to prevent widespread use of transition properties from \dscf{}, and needs to be fixed.

For \dscf{} calculations using Hartree--Fock theory one can clearly proceed by performing nonorthogonal configuration interaction,\cite{Thom2009,Malmqvist1986} not only fixing the transition dipoles but also (presumably) generally improving the quality of the description. On the other hand, for \dscf{} based on DFT, such a procedure is not well defined because the underlying Slater determinants are understood not to be ``the'' wavefunctions, nor the Hamiltonian to be ``the'' Hamiltonian.\cite{Wu2007} It would be possible to build on the approach developed by Wu {et al.}  in the context of constrained DFT, \cite{Wu2007} but that also introduces other choices and approximations.

Here we instead look at the  simple expedient of using symmetric orthogonalization to ensure exact orthogonality.
Recall that symmetric orthogonalization mixes the two states to make a pair of states that are orthogonal while being as close as possible to the original states, and is defined by the transformation
\begin{equation}
    \ket{\Psi_{\tilde\nu}}=\sum_{\nu}\ket{\Psi_\nu}[\mathbf S^{-1/2}]_{\nu\tilde\nu}
\end{equation}
where
$
    \mathbf S=\big(\begin{smallmatrix}
    1&S\\S&1
    \end{smallmatrix}\big)
$
and $S=\braket{\Psi_2|\Psi_1}$.
Based on this transformation, the transition density between the orthogonalized states is given by
\begin{equation}
    \tilde{\mathbf D}^{21}=\frac{1}{4(1+S)}\left[
    (1-a^2)(\mathbf D^{11}+\mathbf D^{22})
    +
    (1+a)^2\mathbf D^{21}
    +
    (1-a)^2\mathbf D^{12} \right] 
    \label{eq:transform_den_mat}
\end{equation}
where $a = \sqrt{1+S}/\sqrt{1-S}$; this parameter is equal to 
1 when \(S = 0\), recovering the expected result $\tilde{\mathbf D}^{21}=\mathbf D^{21}$ in this limit.
In this work we explore the quality of \dscf\ transition dipoles based on the symmetrically orthogonalized transition density.

\section{Computational Details}

Calculations were performed on a set of 109 small closed-shell molecules containing H, C, N, O and F. These structures are a subset of the benchmark set used in reference,\cite{Grimme2016} with molecules of 12 atoms or fewer.

Reference energies and transition dipole moments (reported in atomic units) were calculated for the 3 lowest energy singlet excited states of each molecule using EOM-CCSD with an aug-cc-pVTZ basis set. \cite{Dunning1989a,Prascher2011,Kendall1992} The same quantities were also calculated for the 6 lowest energy singlet excited states using TDDFT with the CAM-B3LYP functional \cite{Yanai2004} and aug-cc-pVTZ basis set. CAM-B3LYP has consistently been shown to perform well for prediction of the optical properties of  both small molecules \cite{Caricato2011,Sarkar2020} and a large number of conjugated chromophores of various sizes. \cite{Grabarek2019,Beerepoot2018,Robinson2018}

Both EOM-CCSD and TDDFT calculations were performed using Gaussian 16.\cite{Gaussian16} Excited states were cross-referenced between the two methods using the symmetry labels provided by Gaussian. In a small number of cases, the symmetry labelling was unsuccessful or defaulted to a different choice (non-abelian or highest order abelian) of point group between the two methods. In these cases, the excited states were matched by hand based on descent in symmetry and their composition, energy, and transition dipole moment. A full list of symmetries and indices of the selected transitions can be found in the supplementary material.

These data were used to benchmark the performance of \dscf{}  in predicting transition properties, both with and without the symmetric orthogonalization correction proposed in equation \ref{eq:transform_den_mat}. \dscf{} calculations were performed in the Entos Qcore package,\cite{qcore} with the CAM-B3LYP functional and aug-cc-pVTZ basis set. 

We investigated only the HOMO-LUMO singlet transition. Using \dscf{}, we calculated the properties of the state corresponding to the spin-conserving excitation of a HOMO electron into the lowest energy virtual orbital.  This does not correspond to a true singlet excitation, which would contain a superposition of $\alpha$ and $\beta$ excitations.  The spin-purification formula,
\begin{equation}
    \Delta E_S = 2\Delta E^{i,\alpha\rightarrow a,\alpha} - \Delta E^{i,\alpha\rightarrow a,\beta},
\end{equation}
was applied to more accurately estimate the true singlet excitation energy. \cite{Ziegler1977,Kowalczyk2013,Hait2020} However, this correction is applied at the end of the SCF cycle and does not affect the composition of the molecular orbitals, which are used to calculate the transition dipole moment. \cite{Hait2020}

Since \dscf{} uses a variational principle to optimise the excited state orbitals, a known weakness is that the calculation can converge on the ground state rather than the desired excited state.  In most cases, this can be prevented using MOM \cite{Gilbert2008}, which selects orbitals to be occupied based on maximum overlap with each occupied molecular orbital in the previous iteration. This stops the orbitals from changing significantly in any particular step of the optimization and helps stabilize the calculation around the excited state stationary points, rather than the global minimum (ground state). However, in a small number of cases, additional help was needed to converge the SCF cycle to the correct excited state.  There are a number of well-established techniques to address this issue.  We used a combination of Fock-damping, modifying the direct inversion of iterative subspace (DIIS) protocol, \cite{Pulay1980,Pulay1982,Hamilton1985} and starting from an initial guess corresponding to excitation of half an electron. 

The properties of the \dscf{} transition were compared to those of the TDDFT transition with the largest coefficient for HOMO-LUMO excitation (based on the orbital indexing in the TDDFT calculation), along with the corresponding EOM-CC transition. For a few molecules this was not an appropriate comparison to make, either because of a reordering of orbitals with very similar energies or because there was no single state dominated by the HOMO-LUMO transition. In these cases, we either selected the correct TDDFT transition by hand or calculated the \dscf{} transition corresponding to the lowest energy TDDFT transition.  Full details of these choices can be found in the supplementary material.

\section{Results}

\begin{figure}
  \includegraphics[width = \textwidth]{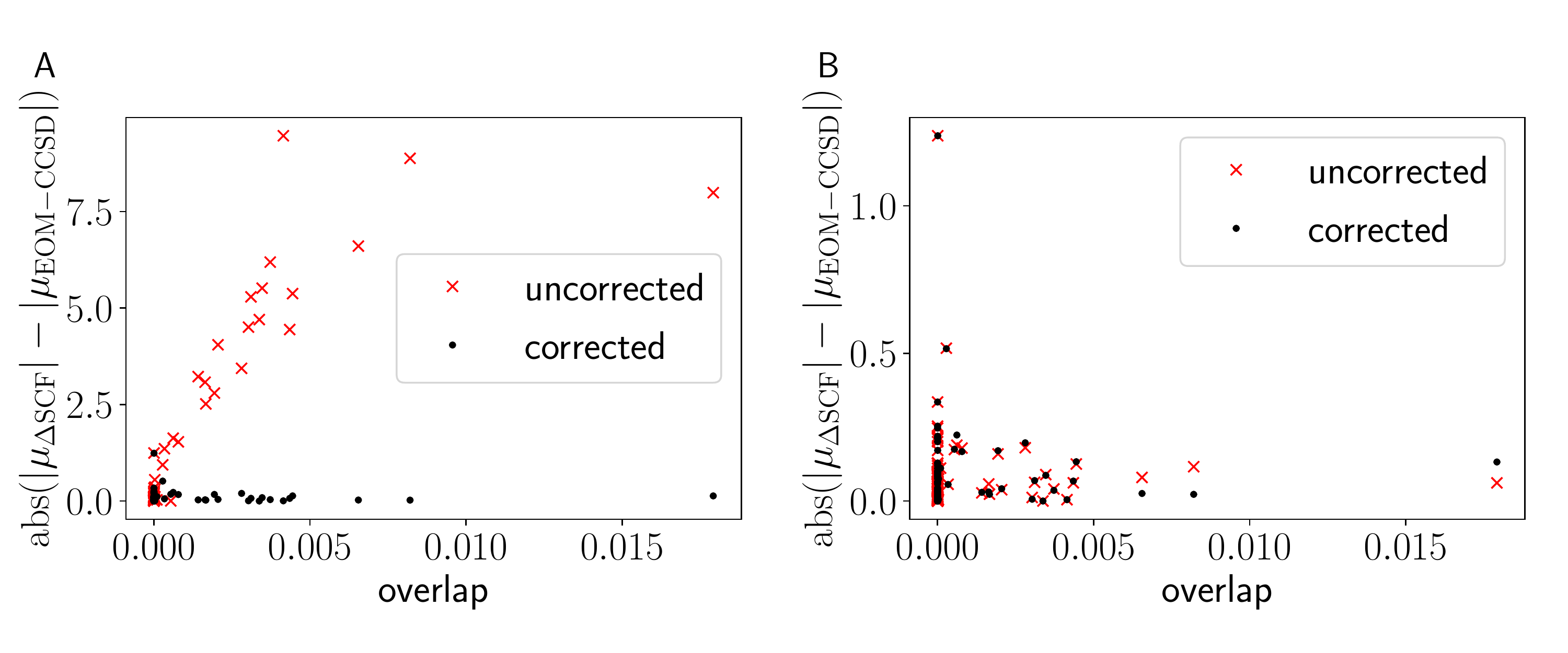}
  \caption{Absolute error in the magnitude of the transition dipole moment calculated using \dscf{} vs. EOM-CCSD, as a function of the overlap between the ground and excited state. The coordinates of the molecules have been translated by \(\left[100, 100, 100\right]\) \AA\ in panel A compared to B. Without any correction (red dots), the coordinate shift results in unphysically large transition dipoles.  This can be avoided by using the symmetric orthogonalization correction (black dots). All calculations used the aug-cc-pVTZ basis set. \dscf{} calculations were performed using the CAM-B3LYP functional.}
  \label{fig:EffectOfCorrection}
\end{figure}

First, we test the effect of applying the symmetric orthogonalization correction, proposed above, to overlapping ground and excited states.  Figure \ref{fig:EffectOfCorrection} shows the error relative to EOM-CCSD in the magnitude of the transition dipole moment, as a function of the ground-excited state overlap for each molecule in the test set using \dscf{} with  or without the correction. In panel A, the coordinates of the entire molecule have been translated by 100 \AA\ in each cartesian direction. Physical properties like excitation energy and transition dipole moment should be invariant under this translation; but when there is non-zero overlap between the ground and excited state, the calculation of the transition dipole moment becomes origin-dependent and this coordinate shift introduces an error into the calculated values of \(|\boldsymbol{\mu}|\).

Although the ground-excited state overlaps are  small (\(<0.2\)) for every molecule in the test set, when the molecule is displaced far away from the origin, it is sufficient to produce highly unphysical transition dipoles.  Using the symmetric orthogonalization correction, the origin dependence is completely removed and these errors do not arise.

An important consideration is whether applying the correction degrades the accuracy of the \dscf{} calculation in any way.  This is difficult to see, since the origin-dependence of the the uncorrected transition dipole moments means that they cannot be taken as a reliable indication of the `correct' \dscf{} transition dipole.  However, we note that, by construction, the amount of ground and excited state dipole that are mixed into the transition density (the amount that the correction `changes the answer') scales roughly linearly with the size of the overlap for small overlaps.  When the overlap is zero (and the uncorrected \dscf{} transition dipole
 is therefore already `correct'), the symmetric orthogonalization procedure does not change the states, transition density or transition dipole at all.  At the largest overlaps present in this test set, the change in the transition dipole that comes from applying the symmetric orthogonalization correction is still very small, as illustrated in panel B of Figure \ref{fig:EffectOfCorrection}.  
 
 Note that we do not attach any significance to whether the corrected or uncorrected transition dipole magnitude is closer to the reference value since the uncorrected magnitude can be made to have any value by shifting the coordinates of the molecule. The molecules in this test set are small, with average atomic positions (not center of mass) 
defining the origin, so we do not expect the uncorrected transition dipole moments to be wildly wrong. However, even shifting the molecule so that its centre-of-mass lies on the origin is sufficient to account for the difference in values seen on the right-hand side of Figure \ref{fig:EffectOfCorrection}.  For the larger molecules in the test set, the transition dipole may not span the whole molecule and the concept of the `correct' position or transition dipole for the molecule becomes even less clear.

For the remainder of this paper, the symmetric orthogonalization correction will be applied for all reported \dscf{} transition dipoles.

\begin{figure}
  \includegraphics[width = \textwidth]{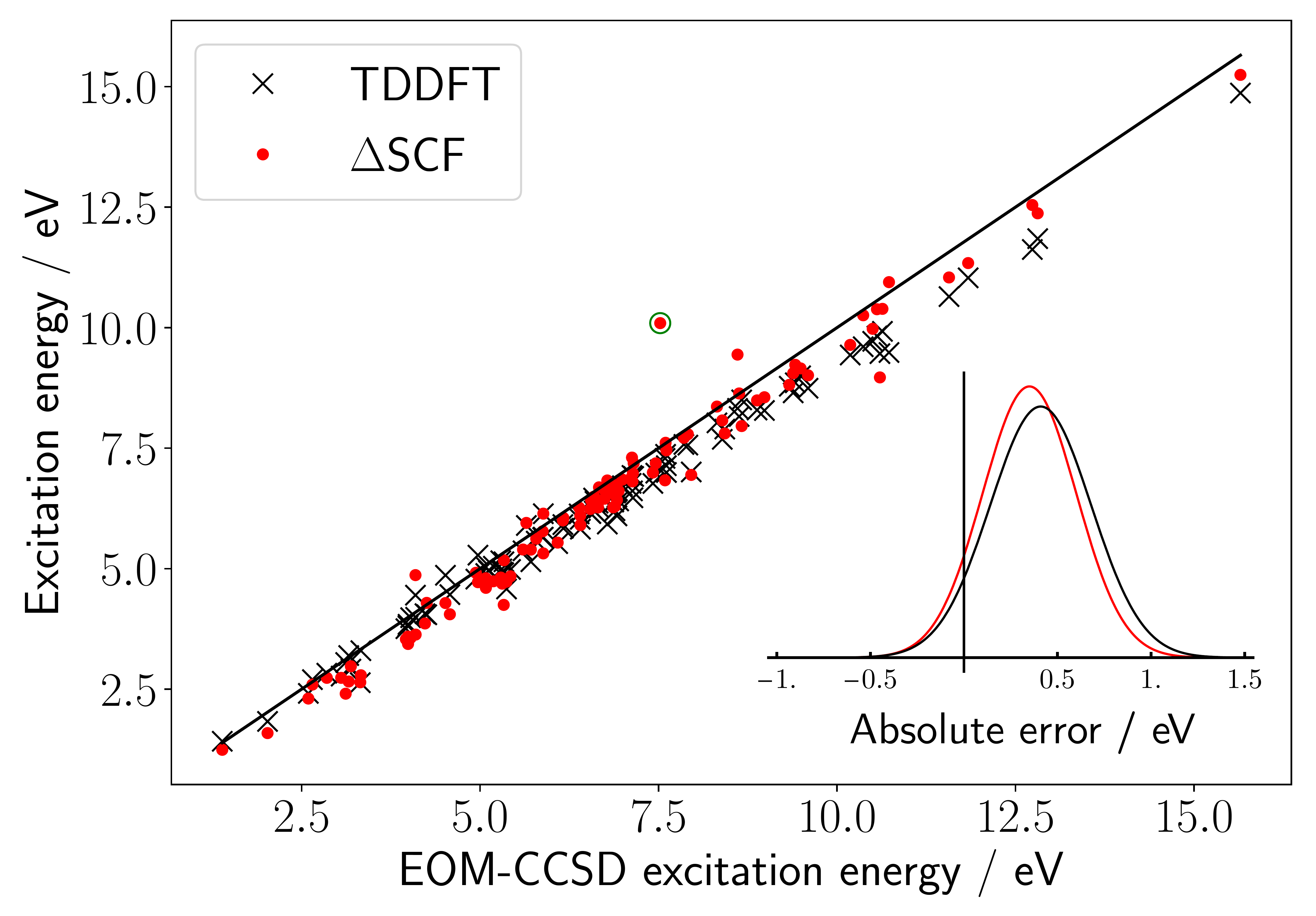}
  \caption{Excitation energies calculated using TDDFT (black cross) or \dscf{} (red dot) compared to the EOM-CCSD reference values. The target values are given by the solid black line \(y=x\). The insert in the lower right hand corner show the probability distribution for the error in each method compared to the EOM-CCSD reference. The outlier circled in green is excluded from this error analysis (see also table \ref{tab:erroranalysis}). All calculations used the aug-cc-pVTZ basis set. \dscf{} and TDDFT calculations were performed using the CAM-B3LYP functional.}
  \label{fig:ExcitationEnergy}
\end{figure}

Figure \ref{fig:ExcitationEnergy} compares the excitation energy of each molecule calculated using TDDFT and \dscf{}  with the value predicted by EOM-CCSD. The energies predicted by \dscf{} are at least as accurate as those predicted using TDDFT, if not slightly more so.  TDDFT with CAM-B3LYP has a tendency to slightly underpredict the excitation energy, which is slightly less pronounced in \dscf{}.  The exception is one very noticeable outlier, highlighted with a circle in figure \ref{fig:ExcitationEnergy}.

This outlier is a perpendicular ethene dimer, and it serves to illustrate a key situation where \dscf{} may not be an appropriate choice of method. The two highest occupied molecular orbitals in the ground state of the ethene dimer are degenerate, representing the \(\pi\)-bonding orbital on each monomer. The two lowest unoccupied molecular orbitals are similarly very close in energy and are in-phase and out-of-phase combinations of the \(\pi\)-antibonding orbitals on each molecule. Both EOM-CCSD and TDDFT predict that the two lowest energy excitations of the ethene dimer are degenerate linear combinations of the local excitations with an excitation energy of 7.5 eV and transition dipole moments in the $x$ and $y$ directions (the principal axis being $z$). \dscf{}, by construction, cannot capture the mixed nature of these excitations, and instead predicts a excitations with  energies around 7 and 10 eV (shown) and transition dipoles in the $xy$ plane.  

Excluding the outlier, the mean  error in the \dscf{}  excitation energies compared to EOM-CCSD is 0.35, with a standard deviation of 0.25 (table \ref{tab:erroranalysis}).  For TDDFT, the mean error is 0.41, with a standard deviation of 0.27. For excitation energies, \dscf{}  is therefore clearly worth considering as a cheap and accurate alternative to TDDFT. This is in good agreement with earlier studies benchmarking \dscf{} excitation energies for large organic dyes.\cite{Kowalczyk2011,Terranova2013}

\begin{figure}
  \includegraphics[width = \textwidth]{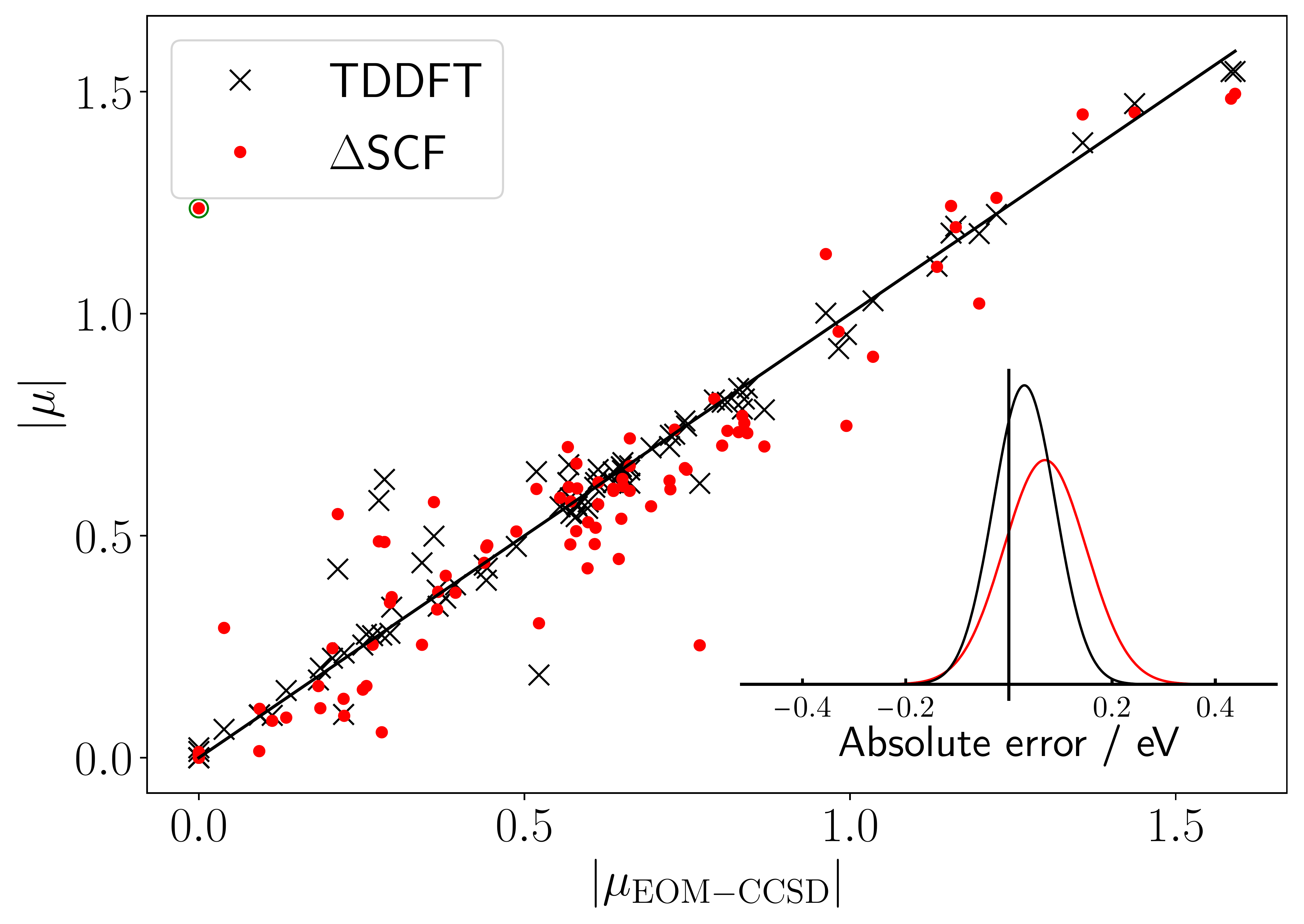}
  \caption{Transition dipole magnitudes (\(|\mathbf{\mu}|\)) calculated using TDDFT (crosses) or \dscf{} (dots) compared to the EOM-CCSD reference values. The target values are given by the solid black line \(y=x\). The insert in the lower right hand corner show the probability distribution for the error in each method compared to the EOM-CCSD reference. The outlier circled in green is excluded from this error analysis (see also table \ref{tab:erroranalysis}). All calculations used the aug-cc-pVTZ basis set. \dscf{} and TDDFT calculations were performed using the CAM-B3LYP functional.}
  \label{fig:TransitionDipoleMagnitude}
\end{figure}

Figure \ref{fig:TransitionDipoleMagnitude} makes the same comparison for \(|\boldsymbol\mu|\). By eye, \dscf{} produces slightly more scatter around the EOM-CCSD reference than TDDFT but has a broadly similar accuracy.  This is borne out in a more detailed numerical analysis.  The mean error in \(|\boldsymbol{\mu}|\) for \dscf{} compared to EOM-CCSD is 0.07, with a standard deviation of 0.08 (table \ref{tab:erroranalysis}).  For TDDFT, the mean error is 0.03, with a standard deviation of 0.06.

There is, again, a single obvious outlier where \dscf{} apparently performs far worse than TDDFT.  This outlier corresponds to a stretched version of the benzene molecule. Like the ethene dimer described above, the lowest energy excitation of this structure is a roughly equal mix of HOMO to LUMO and HOMO$-$1 to LUMO+1 transitions.  In this case, however, both HOMO and HOMO$-$1 and LUMO and LUMO+1 are exactly degenerate and this creates some flexibility in the definition of the transition and its dipole moment. The transition dipole moment found by \dscf{} agrees very well with that for an excitation that is an equal mix of HOMO to LUMO+1 and HOMO$-$1 to LUMO, which, given the degeneracy of the states, is an equally valid choice.  This outlier should therefore be viewed not as a failure of \dscf{} but as a reminder that there isn't one correct transition dipole moment when degenerate states are involved.

This test set contains two other structures for benzene, with slightly different bond lengths. For these variations, coupled cluster and TDDFT find nearly pure, degenerate HOMO to LUMO and HOMO$-$1 to LUMO+1 transitions, for which \dscf{} predicts very accurate transition dipoles.

Having established the performance of \dscf{} vs. TDDFT, we move on to look at the performance of \dscf{} in calculating the transition properties of the 27 BChla in the LHII complex of purple bacteria.  We take the structures of the chromophores from a single snapshot of the molecular dynamics simulation by Stross {\it et al.} \cite{Stross2016}
This chromophore is too large to treat with EOM-CCSD, so we use TDDFT as our reference, bearing in mind its performance on the test set of smaller molecules. We use the PBE0 functional \cite{Adamo1999,Perdew1996} and Def2-SVP basis set \cite{Weigend2005}, in line with Ref. \citenum{Stross2016}.

As shown in Figure \ref{fig:ExcitationEnergy}, the excitation energies calculated using \dscf{} correlate extremely well with those predicted by TDDFT and lie well within the range of error of TDDFT.  This suggests both that the \dscf{} excitation energies are accurate and that small variations in the energy between the different chromophores are physically meaningful.

\begin{figure}
  \includegraphics[width = \textwidth]{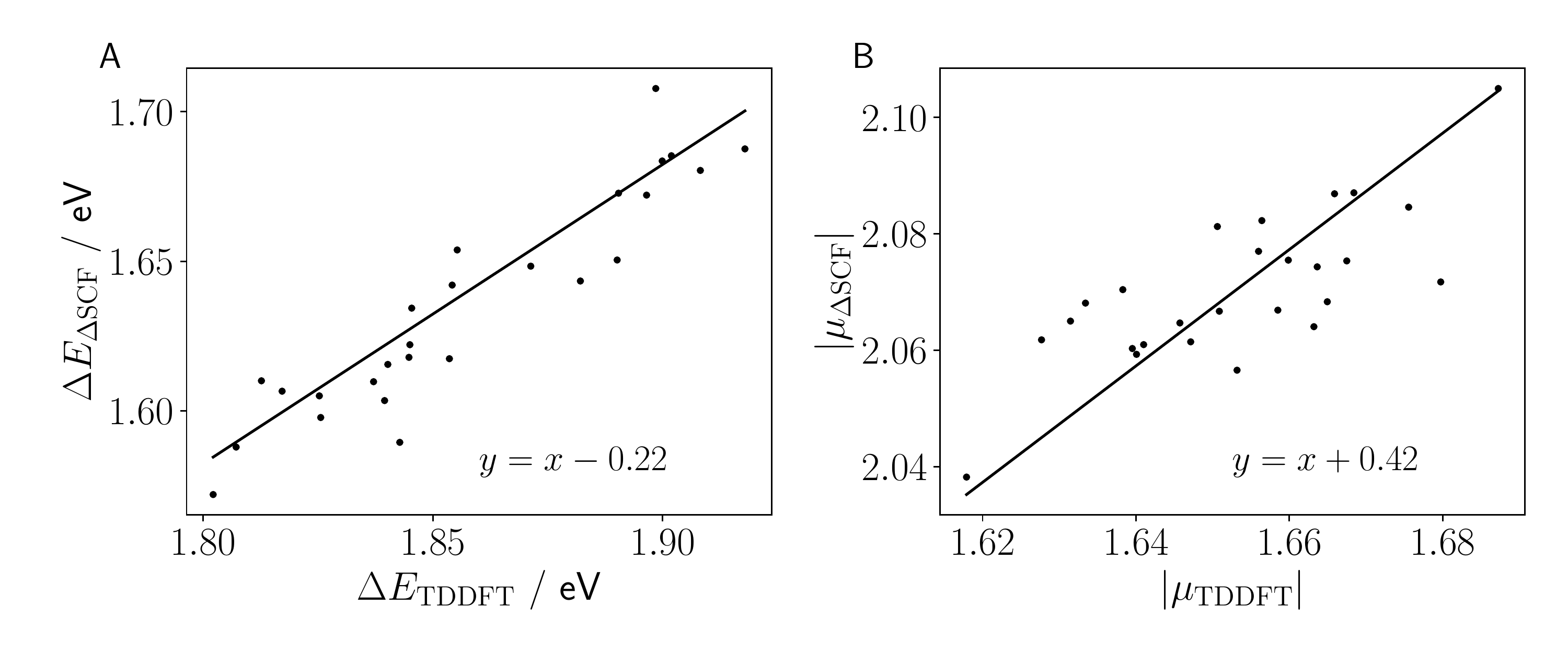}
  \caption{A. Excitation energies and B. transition dipole magnitudes of the 27 BChla molecules in the LHII complex of purple bacteria, calculated using \dscf{} vs. TDDFT. All calculations used the PBE0 functional and Def2-SVP basis set. To highlight the correlation between the methods, we plot the lines \(y = x + C\) on each subplot. The interpretation of the intercept \(C\) is discussed in the text.}
  \label{fig:Chlorophylls}
\end{figure}

By contrast, there is a significant difference between the magnitude of the transition dipoles predicted by TDDFT and \dscf{}, with \dscf{} predicting magnitudes that are, on average, 0.42 a.u. larger.  This is larger than the average error expected for TDDFT and \dscf{} compared to EOM-CCSD but within the full range of errors observed for the test set of small molecules.  We note that the difference between TDDFT and \dscf{} will have contributions from the error in both methods and it is not clear from figure \ref{fig:Chlorophylls} which will be the largest contribution.  However, while it appears that the error in the \dscf{} transition dipole moment is towards the higher end of what we might expect, it is reassuring that the values remain well-correlated with those from TDDFT. This suggests that \dscf{} could be used to create a valid pictures of how the transition dipoles of each chromophore change over the course of a molecular dynamics simulation.

One chromophore is missing from figure \ref{fig:Chlorophylls}, as the \dscf{} calculation collapsed back to the ground state.  This is a hazard of the \dscf{} method, and we plan to keep working on robustness, including for example by implementing the initial maximum overlap method\cite{Barca2018} (IMOM) based on orbitals from an initial averaged calculation. 

\section{Discussion}

We have benchmarked the excitation energies and transition dipole magnitudes predicted by \dscf{} for a large set of small organic molecules. In line with previous work on larger, organic chromophores, we have shown that \dscf{} predicts excitation energies with a very similar accuracy to TDDFT, compared to a highly accurate EOM-CCSD reference. TDDFT still outperforms \dscf{} in predicting the magnitudes of transition dipoles but the error in the \dscf{} predictions are sufficiently small that it can still be considered a useful alternative when TDDFT is 
too computationally demanding or when speed is of greater importance than higher precision.
In contrast to earlier studies, we have focused on testing a large number of different molecules, rather than a range of functionals and basis sets. 

A potential downside of many excited state SCF methods, including the MOM, used here, is that the excited state molecular orbitals are optimised independently of the ground state orbitals and there is consequently no guarantee that the ground and excited state will be orthogonal. In their paper first introducing the MOM, Gilbert {\it et al.} argue that orthogonality is not an expected property of SCF states, which are approximations of the exact quantum states \cite{Gilbert2008}.  They further demonstrate that the MOM tends to converge on excited states that only overlap with the ground state by a small amount. Nevertheless, even a small overlap can introduce a problematic origin dependence into the calculation of the transition dipole moment, particularly when the relevant part of the molecule is not close to the origin of the coordinate axis. We have demonstrated that performing a symmetric orthogonalization of the ground and excited states produced by the SCF optimization is a simple way to remove these small overlaps without introducing error into the calculation of the excitation energy or significantly changing the identity of the states. We have demonstrated the use of this correction in the context of simulating photosynthetic antenna complexes, which consist of multiple chromophores arranged into a larger aggregate structure.  In a molecular dynamics simulation, for example, these complexes would typically be centred globally on the origin, with each individual chromophore therefore being displaced well away from the origin.  Applying our simple correction to the transition density matrix is significantly more  straightforward than recentering every single chromophore (whilst also keeping track of its original position relative to all the other chromophores).  We anticipate that this trick will be extremely useful in the application of cheaper excited-state SCF methods to biological systems.

We have seen that the greatest potential for \dscf{} to fail occurs when the transition of interest is highly mixed in nature.  This is not surprising, since \dscf{} is constructed to deal with transitions between a single occupied ground state orbital and a single (relaxed) virtual orbital.  Highly mixed transitions usually occur when there are low-lying virtual orbitals of the same symmetry with very similar energies. By calculating the energies and symmetries of the molecular orbitals (programs like Gaussian provide an option to do this automatically), a simple inspection would identify molecules with a greater risk of highly mixed transitions, helping to determine whether \dscf{} could be appropriately used. Furthermore, large molecules, for which TDDFT may become prohibitively expensive, typically have much lower symmetry than the small molecules considered here, greatly reducing the chances that near-degenerate molecular orbitals of the same symmetry will exist. 

Looking forward, we suggest that there is potential to further improve the ability of \dscf{} to predict accurate transition dipole moments. Previous work by  Kowalczyk {\it et al.} \cite{Kowalczyk2011} demonstrates that much of the error in \dscf{} excitation energies arises from spin contamination and that this effect is more pronounced for functionals with a smaller amount of exact exchange.  While excitation energies can be, at least partially, corrected for spin contamination using the spin purification formula described above, this correction does not extend to the molecular orbitals used to calculate the transition density, and related properties.  We hypothesise that the performance of \dscf{} for transition dipoles could be improved by incorporating spin purification into the calculation of the molecular orbitals. This could be done, for example, by minimising the spin-purified energy in the SCF cycle, rather than applying the correction at the end of the energy calculation.  Trialling such a procedure is, however, beyond the scope of the current study.


\begin{table}
  \caption{Error in excitation energies and transition dipole magnitudes calculated using \dscf{} and TDDFT at the CAM-B3LYP/aug-cc-pVTZ level of theory.  Errors are calculated relative to an EOM-CCSD reference value. The outliers highlighted in Figures \ref{fig:ExcitationEnergy} and \ref{fig:TransitionDipoleMagnitude}  are excluded from the error analysis for the energies and dipole moment respectively.}
  \label{tab:erroranalysis}
\begin{tabular}{lcccc}
\hline
{} &  \multicolumn{2}{c}{Error in \(\Delta E\)} &  \multicolumn{2}{c}{Error in \(|\bm{\mu}|\)} \\
\hline
 {}&  \dscf{} &  TDDFT  &  \dscf{} &  TDDFT \\
\hline
mean      &                       0.35 &                0.41 &                    0.07 &             0.03 \\
std. dev. &                       0.25 &                0.27 &                    0.08 &             0.06 \\
min.      &                       0.01 &                0.02 &                    0.00 &             0.00 \\
max.      &                       1.63 &                1.24 &                    0.52 &             0.34 \\
\hline
\end{tabular}
\end{table}

\begin{acknowledgments}
We gratefully acknowledge the funding agencies that supported this work: O.F. is funded by the U.S. Department of Energy (DE-FOA-0001912). S.B.W. is supported by a research fellowship from the Royal Commission for the Exhibition of 1851
\end{acknowledgments}

\bibliography{references.bib}

\end{document}